\documentclass[preprint,preprintnumbers,amsmath,amssymb,superscriptaddress]{revtex4}
\usepackage{color}
\usepackage{graphicx}% Include figure files
\usepackage{dcolumn}% Align table columns on decimal point
\usepackage{bm}% bold math
\usepackage{upgreek}
\usepackage{amsmath}
\usepackage{url}
\begin{document}

\title{Highly indistinguishable on-demand resonance fluorescence photons from a deterministic quantum dot micropillar device with 75 $\%$ extraction efficiency}

\author{Sebastian Unsleber}
\affiliation{Technische Physik and Wilhelm Conrad R\"ontgen Research Center for Complex Material Systems, Physikalisches Institut,
Universit\"at W\"urzburg, Am Hubland, D-97074 W\"urzburg, Germany}
\affiliation{These authors contributed equally}
\author{Yu-Ming He} 
\affiliation{Technische Physik and Wilhelm Conrad R\"ontgen Research Center for Complex Material Systems, Physikalisches Institut,
Universit\"at W\"urzburg, Am Hubland, D-97074 W\"urzburg, Germany}
\affiliation{These authors contributed equally}
\affiliation{Hefei National Laboratory for Physical Sciences at the Microscale and Department of Modern Physics,
$\&$ CAS Center for Excellence and Synergetic Innovation Center in Quantum Information and Quantum Physics,
University of Science and Technology of China, Hefei, Anhui 230026, China}

\author{Sebastian Maier} 
\affiliation{Technische Physik and Wilhelm Conrad R\"ontgen Research Center for Complex Material Systems, Physikalisches Institut,
Universit\"at W\"urzburg, Am Hubland, D-97074 W\"urzburg, Germany}
\author{Stefan Gerhardt} 
\affiliation{Technische Physik and Wilhelm Conrad R\"ontgen Research Center for Complex Material Systems, Physikalisches Institut,
Universit\"at W\"urzburg, Am Hubland, D-97074 W\"urzburg, Germany}
\author{Chao-Yang Lu}
\affiliation{Hefei National Laboratory for Physical Sciences at the Microscale and Department of Modern Physics,
$\&$ CAS Center for Excellence and Synergetic Innovation Center in Quantum Information and Quantum Physics,
University of Science and Technology of China, Hefei, Anhui 230026, China}
\author{Jian-Wei Pan}
\affiliation{Hefei National Laboratory for Physical Sciences at the Microscale and Department of Modern Physics,
$\&$ CAS Center for Excellence and Synergetic Innovation Center in Quantum Information and Quantum Physics,
University of Science and Technology of China, Hefei, Anhui 230026, China}
\author{Martin Kamp} 
\affiliation{Technische Physik and Wilhelm Conrad R\"ontgen Research Center for Complex Material Systems, Physikalisches Institut,
Universit\"at W\"urzburg, Am Hubland, D-97074 W\"urzburg, Germany}
\author{Christian Schneider} 
\affiliation{Technische Physik and Wilhelm Conrad R\"ontgen Research Center for Complex Material Systems, Physikalisches Institut,
Universit\"at W\"urzburg, Am Hubland, D-97074 W\"urzburg, Germany}
\email{christian.schneider@physik.uni-wuerzburg.de}
\author{Sven H\"ofling}
\affiliation{Technische Physik and Wilhelm Conrad R\"ontgen Research Center for Complex Material Systems, Physikalisches Institut,
Universit\"at W\"urzburg, Am Hubland, D-97074 W\"urzburg, Germany}
\affiliation{SUPA, School of Physics and Astronomy, University of St Andrews, St Andrews, KY16 9SS, United Kingdom}
\affiliation{Hefei National Laboratory for Physical Sciences at the Microscale and Department of Modern Physics,
$\&$ CAS Center for Excellence and Synergetic Innovation Center in Quantum Information and Quantum Physics,
University of Science and Technology of China, Hefei, Anhui 230026, China}

%\dates{Compiled \today}

%\ociscodes{(230.5590) Quantum-well, -wire and -dot devices; (220.4241) Nanostructure fabrication;  (270.0270) Quantum optics; (300.6470) Spectroscopy, semiconductors}

%\doi{\url{http://dx.doi.org/10.1364/optica.XX.XXXXXX}}

\begin{abstract}
The implementation and engineering of bright and coherent solid state quantum light sources is key for the realization of both on chip and remote quantum networks. Despite tremendous efforts for more than 15 years, the combination of these two key prerequisites in a single, potentially scalable device is a major challenge. Here, we report on the observation of bright and coherent single photon emission generated via pulsed, resonance fluorescence conditions from a single quantum dot (QD) deterministically centered in a micropillar cavity device via cryogenic optical lithography. The brightness of the QD fluorescence is greatly enhanced on resonance with the fundamental mode of the pillar, leading to an overall device efficiency of $\eta=(74\pm4)~\%$ for a single photon emission as pure as $g^{(2)}(0)=0.0092\pm0.0004$. The combination of large Purcell enhancement and resonant pumping conditions allows us to observe a two-photon wave packet overlap up to $\nu=(88\pm3)~\%$.  
\end{abstract}

%\setboolean{displaycopyright}{false}

\maketitle
%\thispagestyle{fancy}
%\ifthenelse{\boolean{shortarticle}}{\abscontent}{}

\section{Introduction}

Building compact and bright solid state sources of on-demand and coherent single photons is a major challenge in engineering quantum emitters and solid state microcavities \cite{michler2009single}. Such sources are of crucial importance for establishing optical quantum networks on- and off chip ~\cite{yao09,Hoang12}, to enable quantum teleportation~\cite{Nilsson-NatPhot13, Gao-NatCom13} and to implement building blocks for quantum repeater networks~\cite{Delteil2015}. Among the most promising candidate for solid state single-photon emitters are semiconductor quantum dots (QDs) grown by epitaxy. Due to their near-unity quantum efficiency and spontaneous emission lifetimes in the lower ns range they allow for operation frequencies in the GHz range. 
As InAs or InP QDs are embedded in a high refractive index medium, improving the extraction efficiency of the single photon flux out of the crystal is a crucial. Thus far, microcavities \cite{Moreau2001,Pelton2002}, lensing \cite{Maier2014, Gschrey2015, Sapienza2015} and waveguiding approaches \cite{Claudon2010a,Heinrich2010,Reimer2012,arcari14} have been most successfully applied to this long standing problem, and single photon sources with extraction efficiencies in excess of 70 $\%$ \cite{Claudon2010a} have been reported. 
The indistinguishability of the emitted photons, however, which is an equally important characteristics of a quantum light source for the above mentioned applications, is a parameter which is harder to achieve. Indistinguishable photons share all spectral characteristics, including polarization and color. Furthermore, the characteristic destructive quantum interference between photons leaving separate output ports on a beamsplitter can only be established if the single photons impinge the splitter at exactly the same time, and if their wave packet overlap equals unity. Equal timing requires an excitation technique which minimizes time jittering of the emission event, while maximizing the wave packet overlap requires close-to Fourier limited photons. While single photon streams can be generated from a single quantum dot under non-resonant or quasi-resonant excitation conditions, it has been recognized that resonance fluorescence excitation is the most suitable configuration to simultaneously minimize time jittering and maximize coherence \cite{He2013,Schneider2015,Unsleber2015b,Ding2015}. 

In this work, we report on a QD micropillar device, for which a preselected QD was placed in the center of a micropillar cavity by a cryogenic insitu lithographiy technique. The resulting device, operated deeply in the Purcell regime, yields an overall device efficiency of $\eta=(74\pm4)~\%$ of resonance fluorescence single photons. The coherence of the resonance fluorescence  photons is measured via Hong-Ou-Mandel type interference experiments resulting in a maximal two-photon wave packet overlap of up to $\nu=(88\pm3)~\%$.  

\section{Technology}

\subsection*{Sample growth and insitu lithography}

\begin{figure}[h]
\centering
\includegraphics[width=0.6\linewidth]{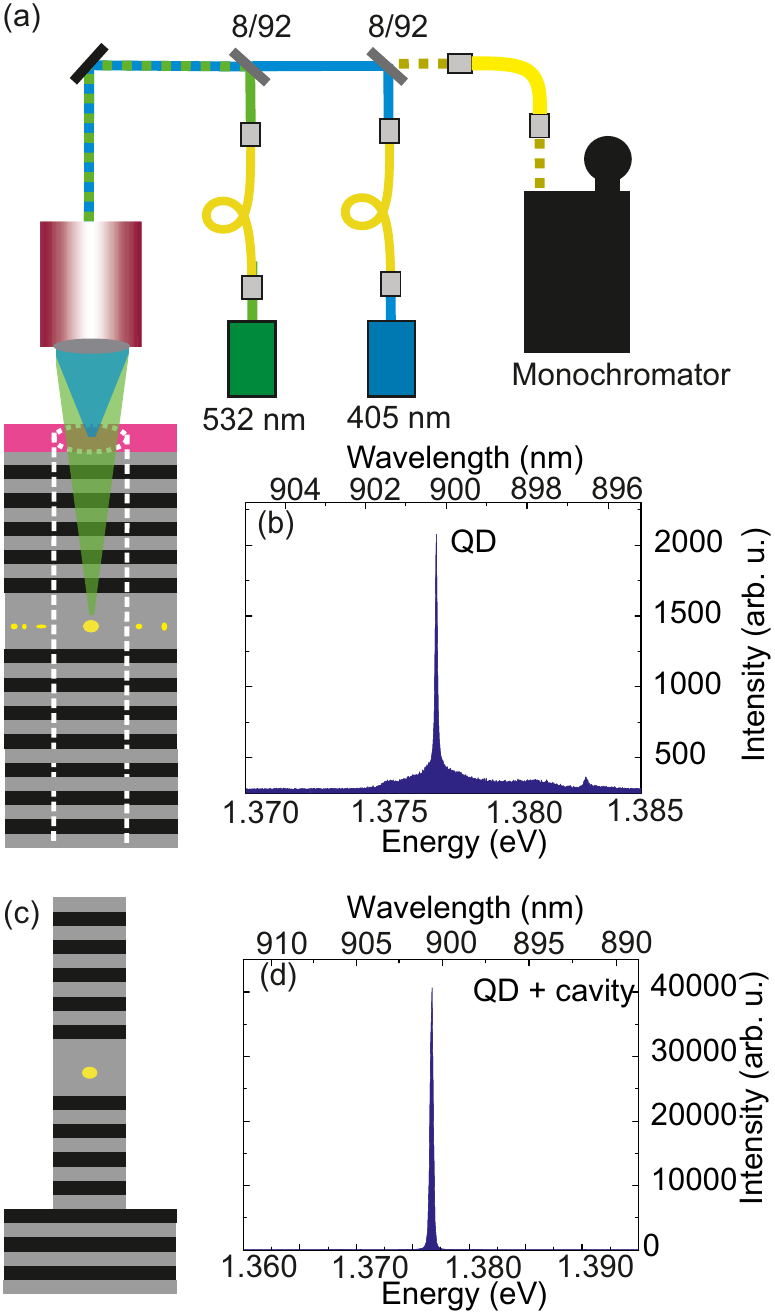}
\caption{Schematic drawing of the deterministic device fabrication: (a) We excite the planar structure with a green laser to pump the QDs without exposing the photo resist. (b) We select a single, bright QD line and optimize the signal. (c) After exposing the photoresist with a violet laser, we etch the micropillar into the planar structure. (d) The result is a bright, background free, Purcell-enhanced single QD line.}
\label{Fig1}
\end{figure}

Figure \ref{Fig1} depicts a sketch of the fabrication scheme which we applied to fabricate our aligned QD-micropillar device, in a similar approach as discussed in \cite{Dousse2008}. The process starts from a planar cavity structure grown via means of molecular beam epitaxy (MBE). It consists of 25.5 (15) $\lambda/4$-thick AlAs/GaAs mirror pairs which form the lower (upper) distributed Bragg reflector (DBR). The DBR mirrors sandwich a $\lambda$-thick GaAs cavity with a single layer of low density In(Ga)As QDs. 
To shift the emission wavelength of the QDs to the 900 nm range, we performed an in-situ partial capping and annealing step while MBE growth.
The sample is spin-coated with a conventional photo resist for optical lithography (sensitive in the ultra violet to blue spectral range) and mounted on the cold finger of a \textit{l}He flow cryostat. We excite the sample with a green, SM-fiber coupled, continuous wave laser ($\lambda=532$ nm) to pump the QDs without exposing the resist. After selecting a bright, resolution limited single QD line (see Fig. \ref{Fig1}(b)), we maximize the signal to ensure that the QD is located in the center of the laser spot. An additional SM-fiber coupled, violet laser ($\lambda=405$ nm), which beam path is identically to the one of the green laser is used to expose the photo resist. Via varying the exposure time, we can control the diameter of the exposed area with sub-micron precision and thus can match the optical resonance to the emission frequency of the QD. Subsequent to this cryogenic optical lithography step, we deposit a BaF/Cr hard mask and after lift-off, the mask pattern is transferred into the sample via electron-cyclotron-resonance reactive-ion-etching. Next, the sample is planarized by a transparent polymer to stabilize the pillars and protect them from sidewall oxidation, and the hard mask is fully removed in an ultra sonic bath.\\
\subsection*{Experimental setup and Hong-Ou-Mandel interference}
The experimental setup for the resonant excitation of such a QD cavity system consits of a linearly polarized Ti:Sapphire laser (repetition rate $82$ MHz, pulse length $\tau\approx1.3$ ps) which is coupled into the beam path via a $92/8$ pellicle beamsplitter. The micropillar sample is mounted on the coldfinger of a \textit{l}He flow cryostat and the emitted light from the sample is collected with a microscope objective (NA$=0.42$) and coupled into a single mode fiber. A second linear polarizer in front of the fiber coupler is orientated perpendicular to the laser polarization and selects the detected polarization axis of the QD signal. After spectrally filtering the emission with a 1500$\frac{\textnormal{lines}}{\textnormal{mm}}$ grating, we either detect the signal on a charged-coupled device(CCD) or we couple it into a fiber. 
The single photon statistics are measured via a fiber-coupled Hanbury Brown and Twiss setup.
For the two-photon interference experiments, we divide each laser pulse into two equal pulses with $2$ ns delay. The interference experiments are carried out with an unbalanced free beam Mach-Zehnder-interferometer, where the two arms exactly compensate the separation between two consecutive emitted photons. We measure the second order auto-correlation of this interferometer via two single photon sensitive Silicon based avalanche photo diodes at the exit ports of the second $50/50$ beamsplitter.

\section{Experimental results}
\begin{figure}[h]
\centering
\includegraphics[width=0.6\linewidth]{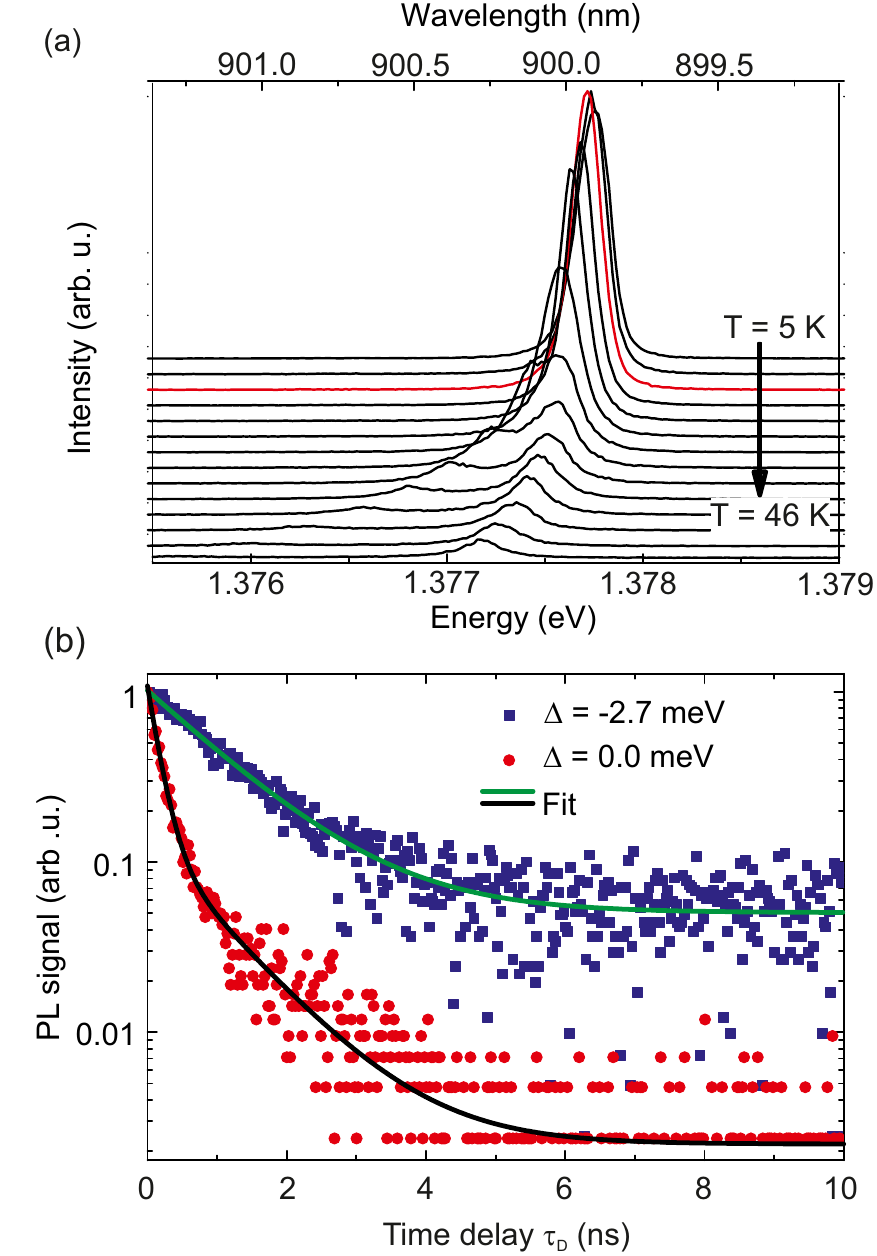}
\caption{(a) Temperature dependent spectras of an insitu defined micropillar with a diameter of $d\approx3~\mu$m under above bandgap excitation. A strong enhancement on spectral resonance due to the Purcell effect is observed. (b) Time resolved measurements on and off resonance reveal a Purcell factor of $F_P=5.8\pm0.2$.}
\label{Fig2}
\end{figure}

In the following, we discuss the emission characteristics of a micropillar with a diameter of $d=3~\mu$m (Q=5930, $\gamma_C=232~\mu$eV). Figure 2(a) shows a series of spectra acquired under non-resonant excitation and for varying temperature. We observe a clear signature of weak light matter coupling with a strong enhancement of the QD emission for spectral resonance between QD and fundamental optical mode. To determine the Purcell enhancement of the system, we carried out time-resolved $\mu$-photoluminescence measurements. The time-depended emission of the QD is plotted in Fig. \ref{Fig2}(b). Via fitting the decay curve for spectral resonance between QD and cavity with a biexponential decay we can extract the lifetime of the QD exciton as $T_{on}=(168\pm5)$ ps. The biexponential behaviour is most likely due to an influence of the dark exction on the PL signal of the QD. We like to mention that this influence is no longer observable for the detuned case, where we had to fit the data with a single exponential decay. The lifetime for a spectral detuning of $\Delta=E_X-E_C=-2.7$ meV$\approx12*\gamma_C$ is measured to be $T_{off}=(1140\pm19)$ ps. This leads to a Purcell factor of $F_P=\frac{T_{off}}{T_{on}}-1=5.8\pm0.2$, under the assumption that suppression of spontaneous emission off resonance is insignificant in our case. This is in perfect agreement with the theoretical maximum \cite{Gerard2002,Barnes2002} of $F_{P,Max.}=\frac{3Q(\lambda/n)^3}{4\pi^2V_M}=5.9$. Here, $n$ is the refractive index of the cavity material ($n=3.6$ for GaAs) and $V_M$ is the mode volume. For our estimation of the maximal Purcell enhancement we used the value reported in \cite{Boeckler2008} and scaled it with the volume of the micropillar.

\begin{figure}[h]
\centering
\includegraphics[width=0.6\linewidth]{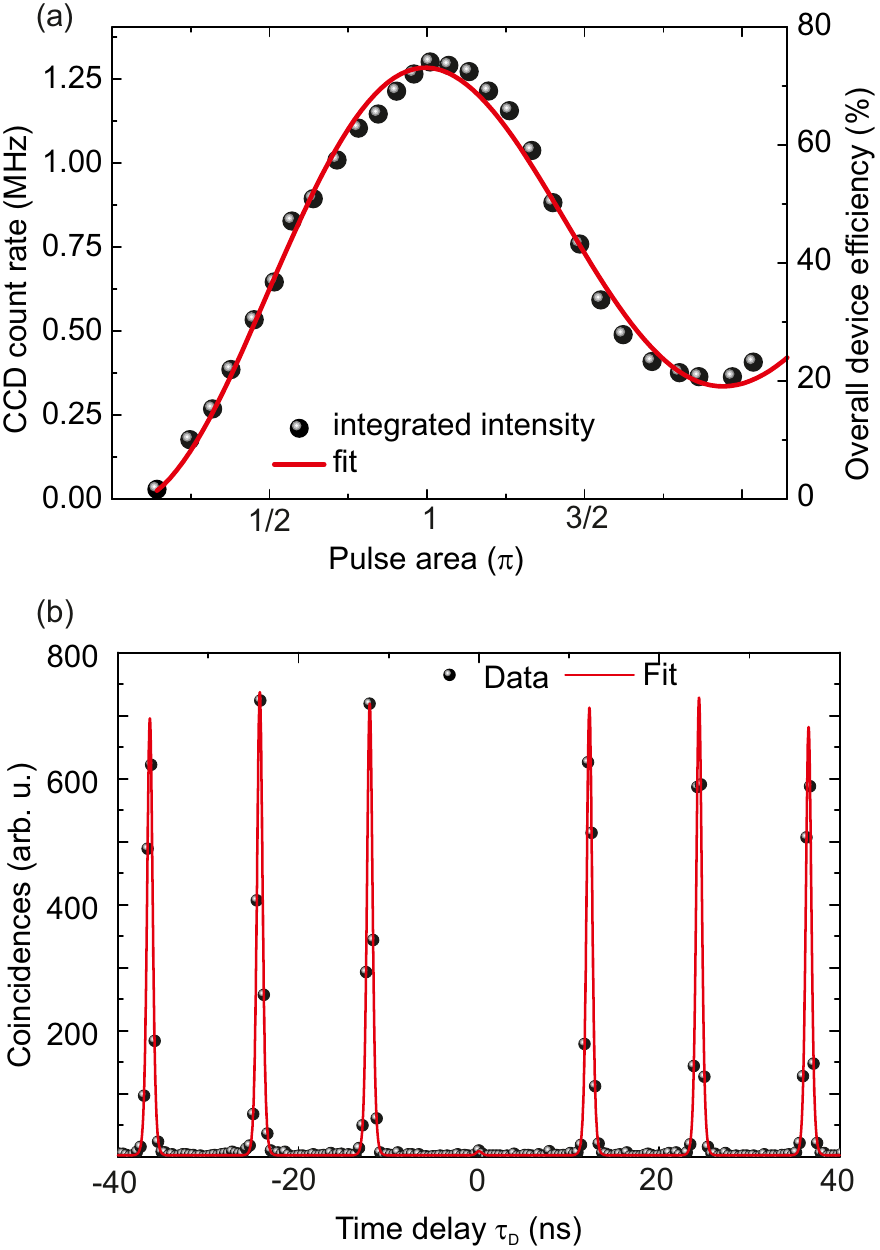}
\caption{(a) Measured source efficiency and count rate on the monochromator CCD versus the pulse area of the driving laser field for spectral resonance between QD and cavity mode. We extract an overall device efficiency of $\eta=(74\pm4)~\%$. (b) 2$^{nd}$ order auto-correlation histogram for pulsed resonant excitation with a $\pi$-pulse. We extract a $g^{(2)}$-value as low as $g^{(2)}(0)=0.0092\pm0.0004$.}
\label{Fig3}
\end{figure}

Full inversion of the two level system can be established most efficiently under strictly resonant pumping: Figure \ref{Fig3}(a) shows the integrated QD emission taken from the CCD of our setup. We observe an oscillating behaviour which is a signature for the pulsed resonant driving of the system. For $\pi$-pulse excitation, we observe count rates on the CCD up to 1.3 million counts per second. In order to extract the overall device efficiency of our QD micropillar device from this measurement, we carefully calibrated our setup revealing a setup efficiency for the detected linear polarization of $\eta_{Setup}=(2.1\pm0.1)~\%$. Therefore, our device yields an overall efficiency of $\eta=(74\pm4)~\%$. This value is comparable to the reported state of the art values for single QD-based light sources. However, we note that it is significantly larger than any other value reported thus far for a deterministically inverted quantum dot via resonant pumping~\cite{Zrenner2002}. The high brightness, and comparably large setup efficiency allows us to record quasi background free single photon correlation charts via a fiber coupled Hanbury Brown and Twiss setup with very good signal to noise ratios on the minute scale. Figure \ref{Fig3}(b) shows the recorded coincidence histogram for $\pi$-pulse excitation. The vanishing peak around $\tau\approx0$ ns is a clear signature of the non classical light emission from the QD. We fit each pulse with a two sided exponential decay convolved with a Gaussian distribution, where the width is the time resolution ($t_{Res}\approx520$ ps) of our setup. This allows us to extract a value of $g^{(2)}(0)$ via dividing the area of the central peak by the average area of the surrounding peaks, which amounts to $g^{(2)}(0)=0.0092\pm0.0004$. These results unambiguously proof pure and bright single photon emission from a deterministically populated exciton state.

\begin{figure}[h]
\centering
\includegraphics[width=\linewidth]{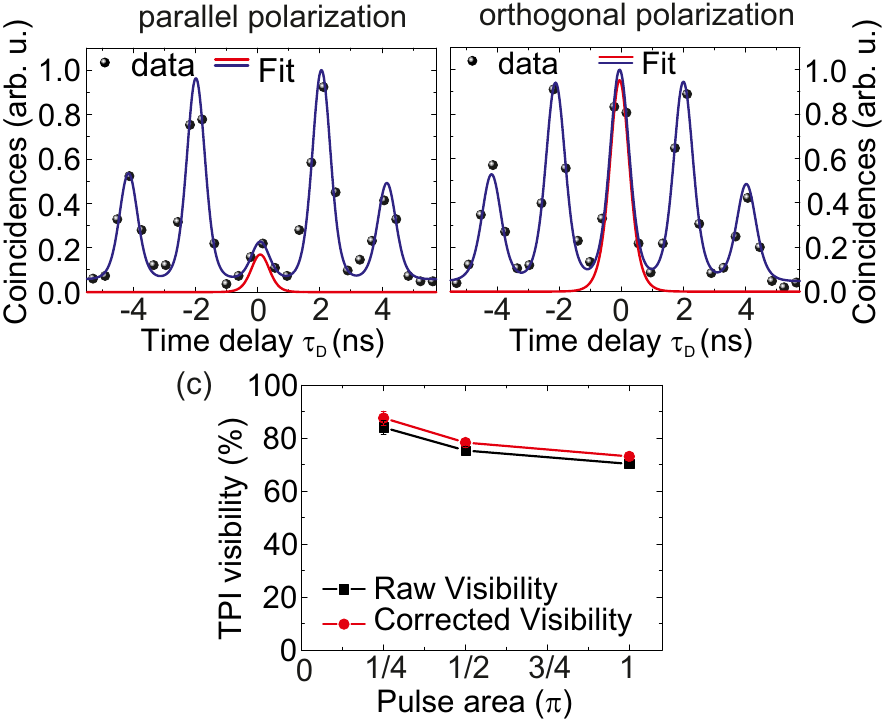}
\caption{Measured two-photon interference histograms for (a) parallel polarization of both photons and (b) orthogonal polarization. We fit each histogram with a sum of five two sided exponential functions each convolved with a Gaussian distribution and extract a visibility of $\nu=(88\pm3)~\%$. (c) For higher pumping strength we observe a modest decrease of the two-photon interference visibility.}
\label{Fig4}
\end{figure}
In order to test the coherence of consecutive emitted single photons from our device, we split the excitation pulses in a train of two pulses with $2$ ns separation. The emitted photons are then coupled into an unbalanced Mach-Zehnder-interferometer. One arm of the interferometer is precisely adjusted to compensate the delay between the two photons. If the early photon takes the long arm of the interferometer and the late photon the short path, both meet each other at the second 50/50 beam splitter where they can interfere if they are indistinguishable. Via an additional $\lambda/2$-wave plate, we can rotate the polarization in one arm of the interferometer by $90^\circ$ to make the two photons distinguishable. Figure \ref{Fig4}(a) and (b) show the measured coincidence histograms for parallel and orthogonal polarization of the photons for driving the system with a $\pi/4$-pulse. The suppression of the central peak in Fig. \ref{Fig4}(a) is a clear proof, that our device is capable as a source for single and highly coherent photons. In order to extract the degree of indistinguishability, we fit each histogram by a sum of five two sided exponential functions, each convoluted with a Gaussian distribution. For the fitting procedure, we keep the time constant of the exponential decay fixed to the measured lifetime of the QD transition (see also Fig. \ref{Fig2}(b)). Via the expression $\nu=1-\frac{A_{parallel}}{A_{orthogonal}}$ we can extract the two-photon interference visibility $\nu$ from our fit results which reveals a raw visibility of $\nu_{raw}=(84\pm3)~\%$. When taking into account the imperfections of the 50/50 beamsplitter (R/T$\approx1.1$), the contrast of the Mach-Zehnder-interferometer ($(1-\epsilon)=0.98$) and the slight deviation from a non zero $g^{(2)}(0)$-value ($g^{*}=0.0092\pm0.0004$), we can use the expressions given in \cite{Santori2002} to correct the two-photon interference visibility which results in a corrected value of $\nu_{corr.}=(88\pm3)~\%$. This number compares favourably to previously reported  values from deterministically fabricated QD based quantum light sources \cite{Gschrey2015,Gazzano2013}. When we increase the pumping strength, we observe a slight decrease of the two-photon wave packet overlap to $\nu=(73\pm1)~\%$ for $\pi$-pulse excitation (see Fig. \ref{Fig4}(c)), most likely as a consequence of power induced dephasing of the system via coupling of the QD to longitudinal acoustic phonons.

\section{Summary}
In conclusion, we have discussed the implementation of a QD based single photon source based on a micropillar aligned to a pre-selected QD via cryogenic optical lithography. Ideal lateral alignment leads to a close to ideal coupling of the QD to the fundamental mode of the micropillar, characterized by a Purcell factor of $F_P=5.8\pm0.2$ which coincides with the theoretical maximum. As a result, we observe bright emission of single photons with an overall efficiency of $\eta=(74\pm4)~\%$ and single photons statistics close to perfection with $g^{(2)}=0.0092\pm0.0004$ under pulsed resonance fluorescence conditions. In a Hong-Ou-Mandel interference experiment, we tested the indistinguishability of the emitted photons revealing a two-photon wave packet overlap as high as $\nu=(88\pm3)~\%$. We believe that our work represents a significant step towards advanced solid state quantum optics experience experiments based on single, indistiguishable photons. Furthermore, the possibility to combine very efficient mode coupling and photon interference in aligned, microstructured devices is very encouraging for the implementation of integrated quantum dot based quantum circuits.

\section*{Funding Information}
 We acknowledge financial support by the State of Bavaria and the German Ministry of Education and Research (BMBF) within the projects Q.com-H and the Chist-era project SSQN.
Y.-M. H. acknowledges support from the Sino-German (CSC-DAAD) Postdoc Scholarship Program.

\section*{Acknowledgements}

The authors would like to thank A. Wolf for performing the etching, planarization and lift-off steps in the lithography scheme.

\end{document}